\documentclass[12pt,english,preprint]{emulateapj}
\usepackage[T1]{fontenc}
\usepackage[latin9]{inputenc}
\usepackage{verbatim}
\usepackage{graphicx}
\usepackage{amssymb}
\usepackage{babel}

\begin{document}

\title{An Emerging Class of Bright, Fast-evolving Supernovae with Low-mass Ejecta}

\author{Hagai B. Perets\altaffilmark{1}, Carles Badenes\altaffilmark{2,3}, Iair Arcavi\altaffilmark{2}, Joshua D. Simon\altaffilmark{4} and Avishay Gal-yam\altaffilmark{2}}

\email{hperets@cfa.harvard.edu}

\altaffiltext{1}{CfA fellow, Harvard-Smithsonian Center for Astrophysics, 60 Garden St., Cambridge, MA 02138, USA}

\altaffiltext{2}{Department of Particle Physics and Astrophysics, Faculty of Physics, The Weizmann Institute of Science, Rehovot 76100, Israel}

\altaffiltext{3}{School of Physics and Astronomy, Tel-Aviv University, Tel-Aviv 69978, Israel}

\altaffiltext{4}{Observatories of the Carnegie Institution of Washington, 813 Santa Barbara St., Pasadena, CA 91101, USA}

\begin{abstract}
A recent analysis of supernova (SN) 2002bj revealed that it was an apparently unique type Ib SN. It showed a high peak luminosity, with absolute magnitude M$_{\rm R}\sim-18.5$,
but an extremely fast-evolving light curve. It had a rise time of $<7$ days followed by a decline
of 0.25 mag per day in B-band, and showed evidence for very low mass of ejecta ($<0.15$ M$_{\odot}$). Here we discuss two additional historical events, SN
1885A and SN 1939B, showing similarly fast light curves and low ejected masses. We discuss the low mass of ejecta inferred from our analysis
of the SN 1885A remnant in M31, and present for the first time the spectrum
of SN 1939B. The old environments of both SN 1885A (in the bulge of M31) and SN
1939B (in an elliptical galaxy with no traces of star
formation activity), strongly support old white dwarf progenitors
for these SNe. We find no clear evidence for helium in the spectrum
of SN 1939B, as might be expected from a helium-shell detonation 
on a white dwarf, suggested to be the origin of SN 2002bj. Finally, 
the discovery of all the observed fast-evolving SNe in nearby galaxies 
suggests that the rate of these peculiar SNe is at least 1-2 \% of all SNe. 
\end{abstract}

\section{Introduction}

\begin{figure}
\includegraphics[scale=0.4]{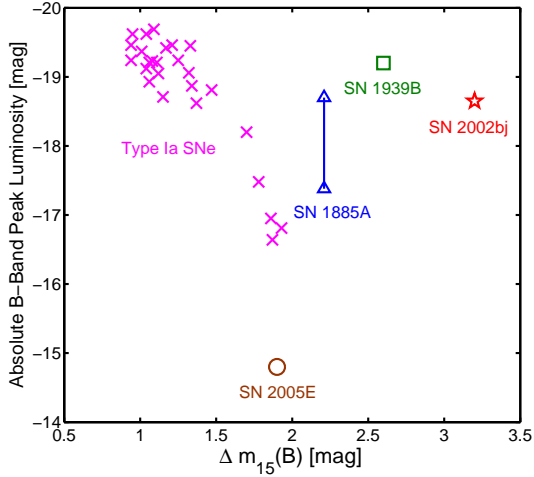}

\caption{\label{fig:b15} $\Delta m_{15}(B)$ for various SNe. Data shown for type Ia SNe ($\times$ signs; taken from \citealt{ben+05b}, table 1), including
faint fast SNe similar to SN 1991bg (e.g. SNe with $\Delta m_{15}(B)>1.7$),
as well as estimated  $\Delta m_{15}(B)$ for SN 2005E, 
interpolated from its light curve ($\circ$ sign; \citealp{per+10}).
 These can be compared with
the  $\Delta m_{15}(B)$ measured for SN 2002bj ($\star$ sign; \citealp{poz+10}) and  SN 1939B ($\square$ sign), and the estimated  $\Delta m_{15}(B)$ for SN 1885A, which has a range of possible peak magnitudes (solid line, connecting the $\triangle$ signs). The specific peak magnitude and $\Delta m_{15}(B)$ values for SNe SN 1885A are unknown, as no direct B-band data are available.  Limiting values for the peak magnitude are obtained under two possible assumptions; (1) a constant $B-V=1.32$ in the 15 days epoch post-peak (bottom triangle), corresponding to the inferred values for SN 1885A assuming a similar evolution as found for sub-luminous SN 1991bg-like SNe \citep{van02,gar+04} and suggested from visual observations \citep{dev+85}, or (2) a constant $B-V\simeq0$ (top triangle) corresponding to a similar evolution to that observed for SN 2002bj. The photographic magnitudes of SN 1939B are taken to be similar to B-magnitudes \citep{bes05}. Error bars are not shown; typical error bars on $\Delta m_{15}(B)$ are $\pm0.1$; typical error bars on peak luminosity are $\pm0.1$, not including error bars from distance modulus (typically $\pm0.15$). Detailed discussion on the error estimates and the $\Delta m_{15}(B)$-peak luminosity relation can be found in \cite{alt+04}. }     
\end{figure}

Recently, various peculiar fast-evolving SNe have been discussed in the literature \citep{per+10,poz+10}.
 \citet{poz+10} studied the type Ib SN 2002bj. They showed its rise time ($<7$ day) and following decay ($\Delta m_{15}(B)\sim3.2$) was much faster than type Ib SNe as well as regular type Ia SNe ($\Delta m_{15}(B)<1.7$; e.g. \citealt{ben+05b}; see Fig. \ref{fig:b15}), and even faster 
than the faint (typically M$_{\rm B}\sim-17$ mag) and fastest SN 1991bg-like events ($1.7<\Delta m_{15}(B)<2$; e.g. \citealt{ben+05b}; see Fig. \ref{fig:b15}). Nevertheless, it reached a peak luminosity of M$_{\rm B}\sim-18.5$ mag, i.e. it was not a faint SN. The total inferred mass in its ejecta was found to be at most $\sim0.15$ M$_{\odot}$. 
\citet{per+10} described a class of type Ib events (with the prototypical SN 2005E), also inferred to have $<1$ M$_{\odot}$ of ejecta, and likely associated with old environments. However, these SNe differ from SN 2002bj as they show faint peak luminosities (typically M$_{\rm B}\sim-15$ mag), they have a slower evolution ($\Delta m_{15}(B)\sim1.9$, although the rise time is unknown) and are calcium rich. 

Such properties for supernovae of type Ib are remarkable. Type Ib
SNe, i.e. showing evidence of helium (but not hydrogen) in their spectra,
are generally thought to originate from the core-collapse explosion
of massive (stripped) stars of at least a few solar masses, and show
much slower evolution. The old environments found for the class of SN 
2005E-like events, and the low ejected mass, found for both types of SNe 
(SN 2005E-like SNe and SN 2002bj),  
therefore suggest different explosion 
model(s) for them. One suggested mechanism is helium-shell 
detonations on white dwarfs  (WDs; \citealp{per+10,poz+10,per+10b,wal+10}).  

The existence of thermonuclear helium-detonation SNe from the explosion of helium-accreting sub-Chandrasekhar mass WDs was theoretically predicted a few decades ago \citep{woo+86,ibe+87,ibe+91,liv+95}.
Variants of these models (including helium deflagration) predicted faint SNe, ejecting  $<1$ M$_{\odot}$ of material (e.g. \citealp{ibe+87,bil+07,she+10,wal+10,woo+10}). 
\citet{bil+07} studied the production of fast evolving SNe with low ejected mass (so called '.Ia' SNe). They initially suggested their light curves evolve even faster than observed for SN 2005E \citep{bil+07}, e.g. $\Delta m_{15}(B)>2$, possibly as fast as observed in SN 2002bj. However, detailed light curve modeling by \citet{she+10}, \citet{wal+10} and \cite{woo+10} show that the expected light curves from their helium-detonation SN models, some of which peak at similar luminosities, decay more slowly than SN 2002bj \citep{wal+10}, while lower luminosity models are possibly similar to SN 2005E, 
under appropriate conditions. 
The rise time of these SNe, however, is comparable with that expected from these models \citep{she+10,wal+10}. 

In the case of SNe similar to SN 2005E, a larger sample (8 SNe) and
several independent properties support the WD helium-shell detonation
scenario (including their old stellar environments and their calcium rich spectra; \citealp{per+10,per+10b}). However, this is not the case for SN 2002bj.

In the following we present the study of two other SNe with bright
but fast-evolving light curves (SNe 1885A and 1939B) similar
to 2002bj in the observed B/V-band. These historical SNe were previously 
suggested to originate from helium-shell
detonations \citep{che+88}, but were not considered 
in the study of SN 2002bj. In this study we would like to
remedy this and give additional data to complement the picture. 

  The data for these SNe are of course less complete than could be obtained with modern instruments.  However, we do not find evidence of any systematic errors or biases, suggesting that the data quality is sufficient to reveal the intrinsic properties of the SNe.

The SNe we discuss provide additional clues to the origin of the fastest
evolving SNe. In particular, they provide direct evidence for old, 
non-star forming environments for at least some of these SNe 
(SNe 1885A and 1939B), likely excluding young massive progenitors 
 and pointing to low-mass, old progenitors such as WDs.
They also provide further evidence for the low ejecta mass (much lower than
the Chandrasekhar mass) of such SNe, and possibly the importance
of other elements and/or processes beside $^{56}$Ni decay in driving the 
SN light curves at early epochs. 
In addition, the low ejecta mass inferred for SN 1885A is
consistent with the low X-ray luminosity of the observed SN remnant
(SNR) in M31, as we will show, thus providing a light-curve-independent
constraint on the SN ejecta. These events also provide us with further data on the frequency of such SNe. Finally, the lack of clear evidence for helium in the spectrum of SN 1939B could give an additional, although puzzling,
clue to the origin of these SNe and the processes involved in their explosions.

\section{Observational data}

\subsection{SN 2002bj}

This recently studied SN was shown to be an unusually fast-evolving 
type Ib SN \citep{poz+10}.
Here we reproduce some of its observations to serve as context for the discussion of the two additional fast-evolving SNe described below.

\subsection{SN 1885A}

\begin{figure}
\includegraphics[scale=0.4]{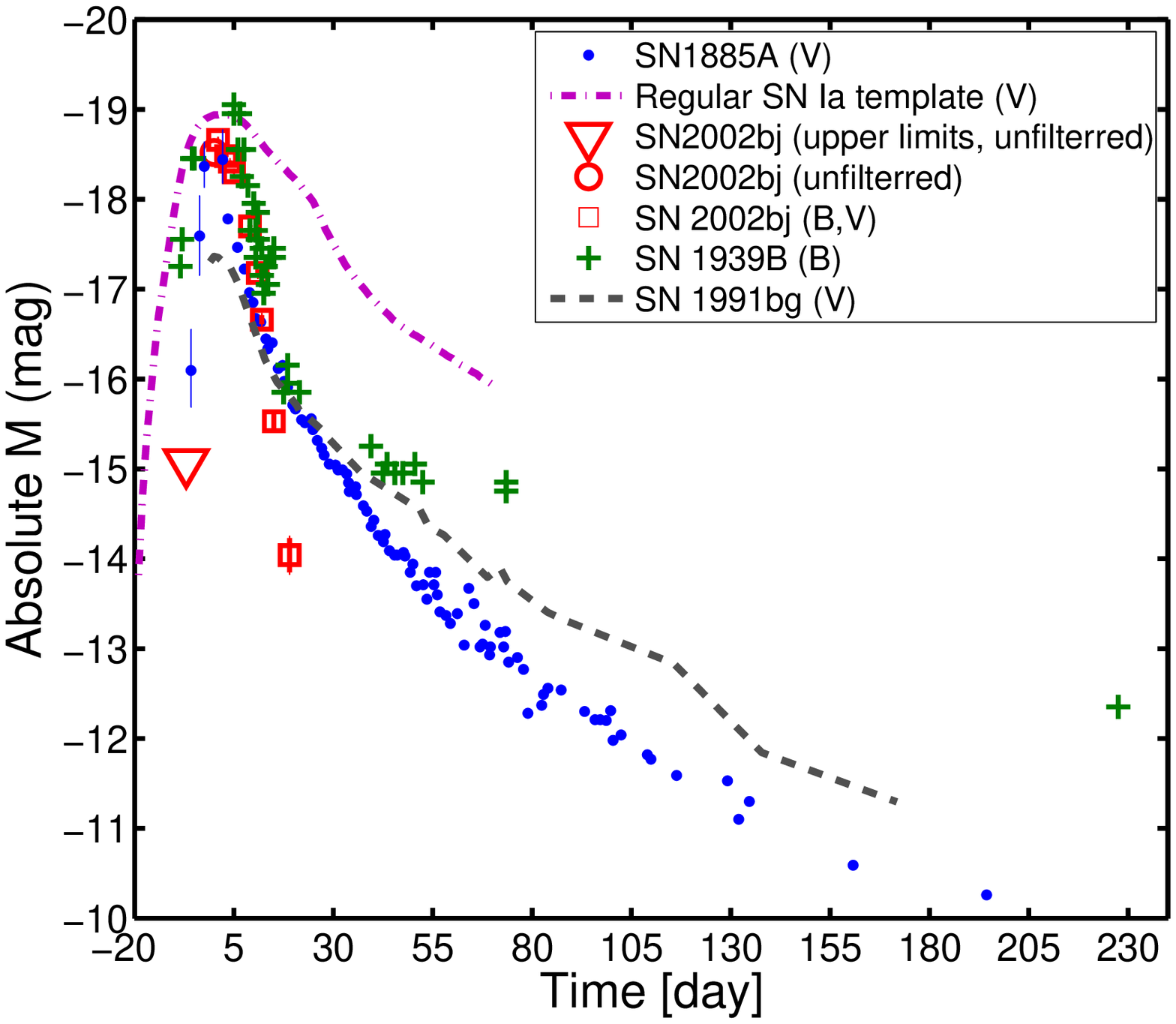} \includegraphics[scale=0.4]{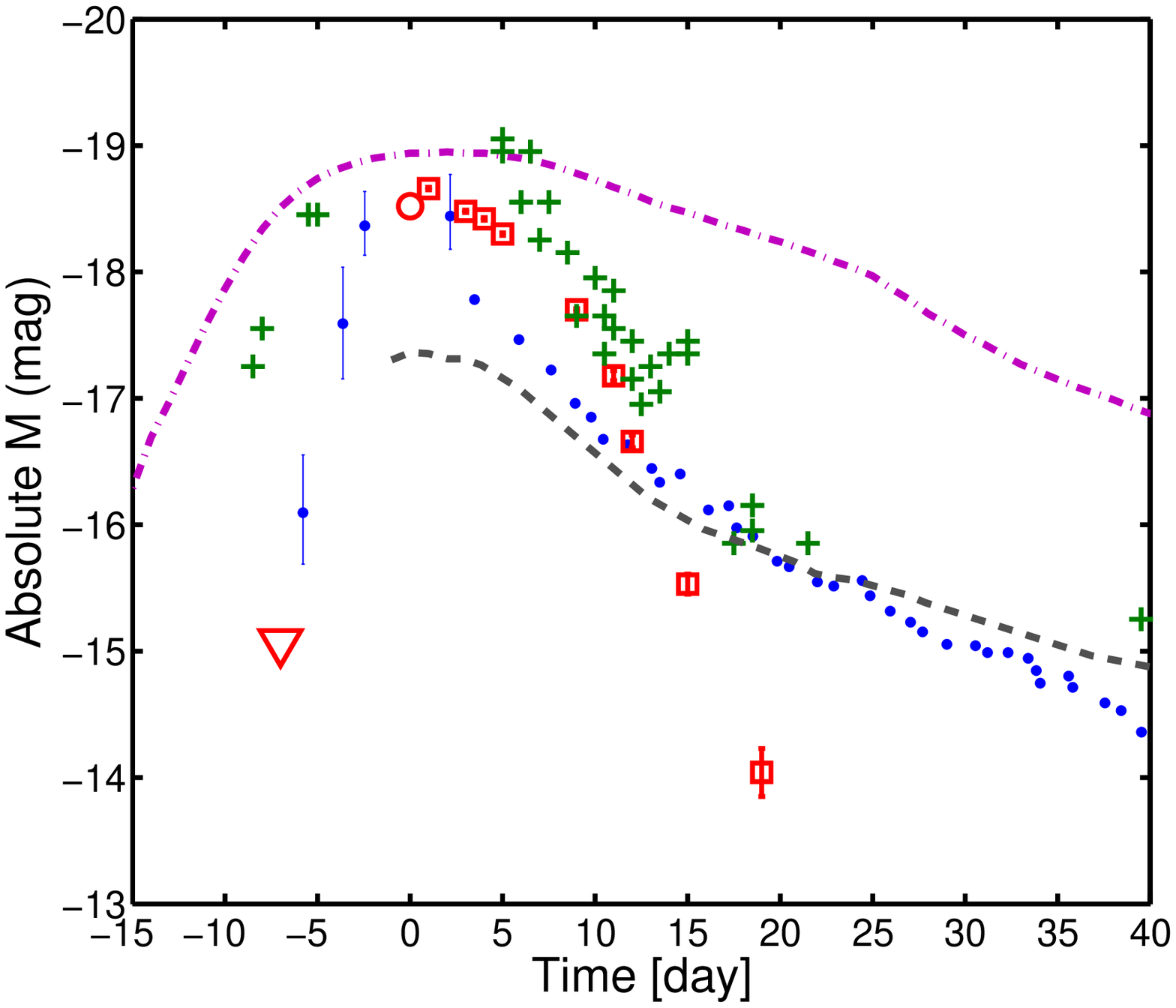}

\caption{\label{fig:The-light-curves}The absolute photographic (pg)/B-band and 
visual (pv)/V-band light curves
of SNe 1885A (pv), 1939B (pg) and 2002bj (B; V band is almost coincident with B). 
Both full light curves (a) and zoomed
in (b) light curves are shown. Note the similar light curve shapes of these 3 events, declining rapidly from a bright peak, in contrast to both normal and faint/fast SNe Ia. The light curves of SNe 1885A and 1939B
are visual light curves, as compiled by \citet[extracted from their published light curve figure]{dev+85}
and \citet{lei+91} for SNe 1885A and 1939B, respectively. The light
curve of SN 2002bj (V band, and unfiltered points) is taken from \citet{poz+10}. Error bars for other data points are unknown,
besides those estimated by \citeauthor{dev+85} for the first data
points of SN 1885A. We show for comparison the V-band light curve of a regular non-stretched SN Ia template \citep{nug+02} normalized to have a peak magnitude of -19. Also shown is the V-band light curve of SN 1991bg (using data from \citealt{lei+93}), the prototypical example for faint and fast evolving type Ia SNe.}
\end{figure}

SN 1885A (S Andromeda), discovered in the bulge of M31, was the first
SN reported in the modern era of astronomy. Its close proximity (the second closest after SN 1987A) enabled many observations
of its light curve to be made following
its discovery in August 1885 (Fig. \ref{fig:The-light-curves}; based on \citealp{dev+85},
data extracted from the published light curve figure) as well as direct observations of its SNR in recent years \citep{fes+89,fes+07}. Unfortunately,
quantitative spectroscopic measurements of SNe did not begin until a few decades after the detection of SN 1885A. Nevertheless, descriptions
of the SN colors, and more importantly several prism measurements,
provide some information on its spectral features.

%

\subsection{SN 1939B}

\begin{figure}
\includegraphics[scale=0.6]{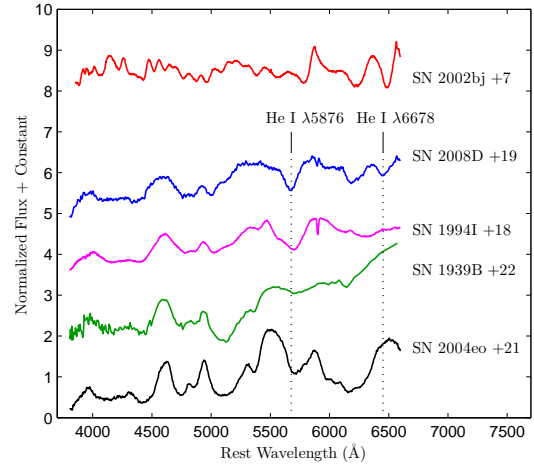}

\caption{\label{fig:Spectrum}The spectrum of SN1939B, 22 days after peak. It is similar to both the type Ia SN 2004eo (from Pastorello et al. 2007) and the type Ic SN 1994I (Filippenko et al. 1995) at a similar phase. The type Ib SN 2008D (from Mazzali et al. 2008) is shown for comparison, with helium lines (blue-shifted by 10,500 km/s) marked. SN 1939B shows no clear evidence for helium, with an apparent absence of the weak He I 6678 absorption. Nevertheless, the feature possibly corresponding to Na D at 5890\AA~+ 5896\AA~ could alternatively be interpreted as a 5876\AA~He line. The spectrum of SN 2002bj at an earlier epoch (7 days) is also shown (from Poznanski et al. 2010; after continuum subtraction).}
\end{figure}

SN 1939B, discovered in the elliptical galaxy NGC 4621 in the Virgo
cluster, was one of the first SNe reported by Zwicky. It was observed
by several people who provided a detailed light curve (Fig. \ref{fig:The-light-curves}, based on data from \citealp{lei+91}, that combined several independent but consistent data sources; re-analysis of some of the original plates by \citealt{sch+99} was also found to be consistent). Although mentioned
in the original discovery report \citep{zwi+39,ada+40}, a spectrum taken 
by Minkowski three weeks post maximum was never published.
With the invaluable help of the Carnegie Observatories 
archive we have been able to locate the original spectrum plate,
and present the reduced spectrum for the first time (Fig. \ref{fig:Spectrum}). In the absence of any documentation, we assume that the observational setup was similar to that described in other contemporaneous works (e.g. \citealt{min39,min42}).
A trace was extracted from both the SN spectrum and the arc spectra by following a line of maximum light. The wavelength solution was fitted using Legendre polynomials, yielding residuals of less than 1 \AA. 
Because no flux calibration is available for the data, the general shape of the spectrum was normalized to most resemble the continuum of SN2004eo at 21 days (Fig. \ref{fig:Spectrum}), enabling a direct comparison of their spectral features.

\section{Properties}

\subsection{Light curves }

The light curves of SNe 1885A, 1939B and 2002bj (as well as SN 1991bg) are shown in Fig.
\ref{fig:The-light-curves}\footnote{Data from the 
NASA/IPAC extragalactic database (NED) were used for distance modulus and Galactic extinction (based on \citealt{sch+98}) of the host galaxies of the SNe; distance modulus error bars are of the order of $\pm0.15$ mag, these are not shown. The light curves are not corrected for extinction inside the host galaxies.}.  
All three light curves peak at luminosities
typical of type Ia SNe, but show a very fast rise to maximum and fast decay. 
The upper limit for the rise time of SN 2002bj is $<7$ days; SN 1885A rises at quite a similar timescale of $\sim6$ days. The rise time of SN 1939B is somewhat longer, $\sim7-8$ days.  All three SNe decay fast, faster than SN 1991bg-like sub-luminous SNe, with  $\Delta m_{15}(B)$ of $\sim2.2,\,2.6$ and $\sim3.2$ for SNe 1885A, 1939B and 2002bj, respectively, where we assume a $\sim$constant $B-V$ for SN 1885A during the 15 days post-peak 
(as seen for SN 2002bj, and inferred for SN 1885A, e.g. \citealt{dev+85,van02}), as only V-band data are available for this SN. 

The light curves of SNe 1885A and 1939B were followed for
a long period and show a break at $\sim60$ and $\sim20$ days, respectively,
with a power-law-like decay afterward\footnote{Note that some photographic light curves of SNe show a leveling at late epochs, possibly due to background contamination \citep{lei+91}. However, when discussing this effect Leibundgut et al. (1991) specify several such leveling SNe, but do not consider SN 1939B to be one of these. Moreover, SN 1939B does not level out to some possible background level as might be expected by an interpretation of background light [as indeed seen in the leveling of SNe specifically mentioned by Leibundgut et al. (1991)  in that context], but continues to fainter magnitudes, e.g. the latest point of  SN 1939B is more than two magnitudes fainter than the position of the light curve break. Similarly, SN 1885A does not show signs of leveling, and continuously becomes much fainter at late times after the break.}. 
The slope of SN 1885A at this epoch ($0.028-0.029$ mag per day) is similar to that observed for type Ia SNe at 
late epochs (see the non-stretched SN Ia template light curve in Fig. \ref{fig:The-light-curves}, and the light curve of SN 1991bg). The light curve of SN 1939B, however, decays more slowly after the break ($0.022-0.024$ mag per day).

The late time exponential decay behavior of SN 1885A may suggest a similar origin as for regular type Ia SNe at this epoch, namely the radioactive decay of $^{56}$Co. The resemblance between SNe 1885A and 1991bg at this epoch is in stark contrast to their early behavior, showing a large difference between their peak luminosities. In type Ia SNe the peak luminosity is thought to be driven by the energy deposition from $^{56}$Ni decay, whereas the late decay is thought to be driven by its product ($^{56}$Co). At early epochs the $\gamma$-rays (emitted as $^{56}$Ni decays to $^{56}$Co and then to $^{56}$Fe) are still trapped in the ejecta and deposit their energy there through successive Compton scatterings. 
At later times, however, and especially for the case of low ejecta mass (as is likely the case for the fast evolving SNe) 
the column density of the ejecta may may not be 
thick enough to trap the $\gamma$-rays and they may escape without depositing
their energy into the ejecta. In principle, this could explain the difference between the luminosities of the SNe at different epochs, as SN 1885A seems to have much lower mass of ejecta. However, when estimating the difference in the fraction of $\gamma$-rays absorbed inside the ejecta we find that it could only account for a small fraction of the observed difference, as we now show.

The electron (Thompson) optical depth is: 
\begin{eqnarray}
\tau_{e} & =& n_e R \sigma_{T}  = \frac{3}{4\pi} \frac{M_{\rm ej}}{m_p} \frac{Z}{A} \frac{\sigma_{T}}{R^2} \cr
             & \sim& 9\left(\frac{M_{\rm ej}}{0.16 M_{\odot}}\right) \left(\frac{t}{15\,{\rm day}}\right)^{-2} 
\end{eqnarray}
where $Z$ is the atomic number, $A$ is the mass number, m$_p$ is
mass of proton, $\sigma_T$ is the Thompson cross-section and $R\sim
{\rm 6(t/day)\,AU}$ is the radius at time $t$. Therefore the optical depth 
at the epoch of peak luminosity is sufficient to trap most of 
the $\gamma$-rays. Assuming a similar energy source from radioactive decay, 
we expect the ratio between the luminosities of SNe 1885A and SN 1991bg to reflect the ratio between their $^{56}$Ni ejecta mass, M$_{\rm Ni}$.   
 Next, following \citet{kas+10} we use a fitting formula 
(as given in \citealt{kul05}; Equation 47) to estimate the 
fraction of $\gamma$-rays that are effectively
absorbed inside the ejecta, $\eta(\tau_e)$. At late times (100 days)
we find $\eta$ to be $\sim0.011$ M$_{\odot}$ and $\sim0.057$ M$_{\odot}$ for SN ejecta of $0.2$ M$_{\odot}$ (corresponding to the mass estimate derived below for SN 1885A) and $1.4$ M$_\odot$ (for a typical type Ia SN), respectively. 
At this epoch the radiation is dominated by the energy deposition from the 
radioactive decay of $^{56}$Co rather than the  $^{56}$Ni decay 
(which has a half life time of only $\sim6$ days), and therefore 
${\rm L_{rad}}\simeq(0.965\eta + 0.035) {\rm L_{Co}}$ 
where L$_{\rm Co}$ is the radioactive power released by the decay of $^{56}$Co (following \citealt{kas+10}, where the kinetic energy of positrons is 3.5 \% of L$_{\rm Co}$; \citealt{sol+02}). We find that the difference in $\gamma$-rays absorption between the two SNe changes their relative luminosities by only a factor of  $(0.965\eta(1885A) + 0.035)/(0.965\eta({\rm1991bg}) + 0.035)\simeq0.045/0.09=0.5$. As can be seen in Fig. \ref{fig:The-light-curves} the ratio between the SN luminosities at peak vs. late epoch can not be accounted for by this factor. Even if the effective optical depth is ten times larger than the simple Thompson optical depth we use (e.g. \citealt{swa+95} find few to ten times larger depth) the ratio we find changes only by order unity (0.4 instead of 0.5). 
We note that the estimated Thompson optical depth we use is calculated at 
the bottom of the ejecta, where as not all of the $\gamma$-rays is 
emitted from there. It is therefore 
possible that a large fraction of the $\gamma$-rays may not be trapped 
even at peak luminosity. This issue is more likely to affect the less massive 
ejecta of SN 1939B, in which case the observed peak luminosity of SN 1939B 
is even lower than it would have been if all the $\gamma$-rays were trapped.
The inconsistency between the peak and the  tail ratios for SN 1939B and SN 1991bg which we discuss might therefore be even larger than we calculated.
We conclude that one requires a different/additional mechanism to drive the light curve at an early epoch. One possible explanation is that SN 1885A had other radioactive elements with short lifetime which contributed to the early light curve generation (e.g. \citealt{she+10,wal+10}). 

Note that an additional data point at day 20 (post-maximum) exists
for SN 2002bj in bands other than V (see figure 1 in \citealp{poz+10}), 
possibly suggesting a steeper slope
than SN 1885A and SN 1939B at this epoch. This is seen in the B and R bands, which likely bracket the V band to a similar magnitude. The I band behavior, however, shows a less steep slope, and is quite unique; we see an $R-I\simeq1.2$ for SN 2002bj at this epoch, which is much larger than found for any type of SNe (compare with \citealp{poz+02}).
The color evolution of SN 1885A and SN 1939B is unknown, as no filtered data are available. Color descriptions summarized by \citet{dev+85} suggest a value of $B-V\sim1.3$ for SN 1885A at peak, much larger than the $B-V\sim0$ observed for SN 2002bj.  
\subsection{SN 1885A remnant\label{sub:Supernova-remnant}}

SN 1885A is one of the very rare cases where both the SN and its SNR
could be observed. After several failed attempts to find optical,
radio, and X-ray emission at the location of SN 1885A, the SNR was finally
discovered \textit{in absorption} against the bulge of M31 \citep{fes+89}
and later studied in detail using HST \citep{fes+07}. The lack of
direct emission from this 125 yr old SNR is remarkable. More than
a decade of monitoring X-ray binaries in the central region of M31
with \textit{Chandra} \citep{wil+06,li+09} has only yielded an upper
limit on the X-ray luminosity of $\sim6\times10^{34}$ erg s$^{-1}$ from the SNR,
assuming a Raymond-Smith model with k$_{b}$T$=0.4\,$KeV (Zhiyuan Li, private
communication, 2010).

\begin{figure}
\includegraphics[scale=0.9,angle=90,origin=c]{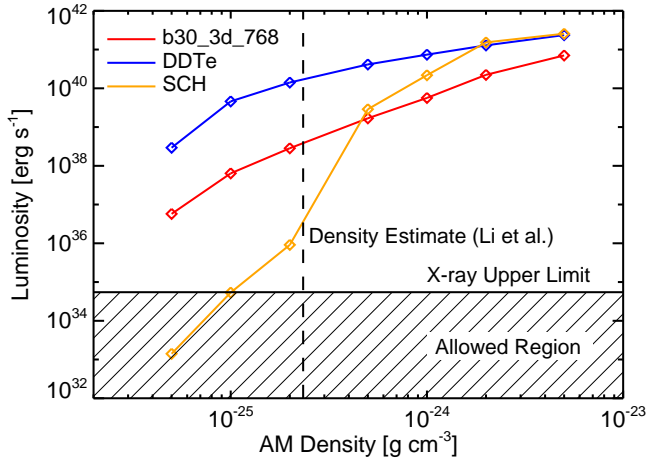}

\caption{\label{fig:SNR}Predicted thermal X-ray luminosities from the shocked
ejecta in several SNR models at an age of 125 yr, plotted as a function
of ambient medium density. The models are: b30\_3d\_768 (from \citealp{tra+04};
M$_{\rm ej}=1.4$ M$_{\odot}$; E$_{k}=0.5\times10^{51}$ ergs; red line),
delayed-detonation (DDTe; from \citealp{bad+03}; M$_{\rm ej}=1.4$ M$_{\odot}$; E$_{k}=1.0\times10^{51}$
ergs; blue line), and sub-Chandrasekhar (SCH; also from \citealp{bad+03}; M$_{\rm ej}=1.0$ M$_{\odot}$;
E$_{k}=1.0\times10^{51}$ ergs; orange line). The horizontal line
is the upper limit on the X-ray luminosity of the SNR (Zhiyuan Li,
private communication, 2010). The vertical dashed line is the estimate
of the density in the circumnuclear region of M31, which includes
the location of SN 1885A from \citet{li+09}. }

\end{figure}

The thermal X-ray emission from a young SNR is determined mainly by
the density of the ambient medium and the mass and kinetic energy
of the SN ejecta (e.g., \citealp{ham+84,bad+03}). \citet{li+09}
have analyzed the properties of the diffuse X-ray emission in the
central arcminute of M31 (which includes the position of SN 1885A),
and find an average density of $\sim0.1$ cm$^{-3}$. This gas density
is low, as befits the bulge of a spiral galaxy, but only an order
of magnitude lower than the ambient medium density around the Tycho SNR \citep{bad+06},
and only a factor 2-3 lower than around the remnant of SN 1006 \citep{ray+07}, which are prototypical Type Ia SNRs. We used
the code described in \citet{bad+03,bad+05} to calculate the thermal
X-ray luminosity of SNR models with an age of $125$ yrs, assuming
different ambient medium densities and SN ejecta configurations. For
all canonical Type Ia SN ejecta (i.e., explosions with M$_{\rm ejecta}\sim1.4$ M$_{\odot}\simeq$ M$_{\rm Ch}$
and kinetic energy of at least a few $\times10^{50}$ erg; see Fig.
\ref{fig:SNR} for the specific models used), our calculations predict
X-ray luminosities in excess of $10^{38}$ ergs s$^{-1}$ at the value
of the ambient medium density estimated by \citet{li+09}, and well above $10^{37}$
ergs s$^{-1}$ even for densities a factor two lower. Such high luminosities
are inconsistent with the observed X-ray upper limit, even after correcting
for absorption. Only Type Ia SN models with significantly lower ejected
masses, such as sub-Chandrasekhar explosions, might be able to reproduce
X-ray fluxes low enough to match the observations. It is of course possible that SN 1885A is
in a local 'bubble' with a density much lower than the average value
found by \citet{li+09}, but taken at face value, the lack of direct
emission from the remnant of SN 1885A seems hard to reconcile with the possibility
that its birth event was a canonical Type Ia SN, and suggests much
lower ejecta mass and energy.

\subsection{Spectra }

The available spectra of the fast-evolving SNe are shown in Fig. \ref{fig:Spectrum}.
A good quantitative spectrum of SN 1885A is not available.
 A historical reconstruction based on observations made with prisms 
is described by \citet{dev+85}. The spectral data suggest that it is not a type II SN, but given its poor quality, it cannot be used to conclusively classify SN 1885A as a specific type I SN. We do note that
\citet{dev+85} and  \citet{pas+08} mentioned 
the possibility
of a type Ib/c identification (the latter suggesting SN 1885A was a massive star exploding in a He-rich circumstellar medium, though the stellar environment of this SN is likely old). However, detailed analysis by \citet{whe+90} argue against any evidence for helium lines. We conclude that SN 1885A most resembles a type Ia or a type Ic SN.

The spectrum of SN 1939B, taken $\sim22$ days after peak, seems to 
resemble both that of regular type Ia and type Ic SNe, e.g. SN 2004eo (Ia) and SN 1994I (Ic; see fig. \ref{fig:Spectrum} 
for comparison) as well as faint SNe such as SN 1986G (spectrum not shown). 
We find no evidence of hydrogen, and no clear evidence for helium lines in the spectrum. 

Naturally, the recently studied SN 2002bj has much better spectral
observations than the historical SNe. Possible helium lines are seen 
in its spectra and may show it to be a type Ib SN, however, \cite{kas+10} 
suggest a different interpretation, making this SN more similar to a type
Ia or a type Ic SN. Note also that this SN shows other similarities 
with type Ia SNe, mostly in its absorption feature near 6150 \AA~
(rest frame), usually attributed to Si II \citep{poz+10}.

\subsection{Inferred masses}
Calculations based on the fast light curve of 
SN 2002bj \citep{poz+10}, resulted in $\sim0.15$ M$_{\odot}$ of mass in the ejecta. 
We can use a similar treatment, following \citet{arn82}, by relating the mass, kinetic energy, and the timescale of the SN light curve, to estimate the ejecta mass in SN 1939B. 
The ejecta mass of a given supernova can be estimated using its light curve and the
observed ejecta velocities. 
The expansion velocity, $v$, of a SN is proportional to $({\rm E_{kin}}/{\rm M_{ej}})^{1/2}$, where E$_{\rm kin}$ is the kinetic energy and 
M$_{\rm ej}$ is the ejected mass, while the typical duration of a SN light 
curve is t$_{\rm d}\propto ($M$_{\rm ej}^{3} /$E$_{\rm kin})^{1/4}$ \citep{arn82}.  
Combining these equations and
assuming that two objects have the same opacity, we have
\begin{equation}
  E_{{\rm kin, } 1} / E_{{\rm kin, } 2} = \left ( \frac{v_{1}}{v_{2}}
    \right )^{3} \left ( \frac{t_{1}}{t_{2}} \right )^{2}
\end{equation}
and
\begin{equation}
  M_{{\rm ej}, 1} / M_{{\rm ej, } 2} = \frac{v_{1}}{v_{2}} \left (
    \frac{t_{1}}{t_{2}} \right )^{2}.
\end{equation}
\noindent

Given the
spectral similarity of SN 1939B (line widths) to SNe 1986G and 2004eo, 
we take 
the photospheric velocities to be similar to these. We adopt a velocity of $10^4$ km s$^{-1}$ (given the similarity to SN 1986G and SN 2004eo, which were found to have typical SN Ia photospheric velocities of $9.36\times10^3$ and $10.52\times10^3$ km s$^{-1}$, respectively). The timescale
for the rise and fall of SN 1939B is comparable to that of SN 2002bj, although slightly longer (where \citealt{poz+10} find the rise time to be 3 times shorter than typical SNe Ia). Taking a rise time of $t_{rise}\sim7$ days, and comparing these values with an average type Ia SN with $t_{Ia-rise}\sim17.5$ days, and taking the following typical velocities of $10^4$ km s$^{-1}$ \citep{hay+10}, total ejecta mass of $1.4$ M$_\odot$ and total energy of $10^{51}$ erg, we find the ejecta mass of SN 1939B to be $\sim0.22$ M$_\odot$ with a kinetic energy of 
$1.6\times10^{50}$ (where similar opacities are assumed). No data are available for the photospheric velocities of SN 1885A and similar direct estimate can not be done. However, if we take the expansion velocity of its SNR, $1.24\pm0.14\times10^4$ km s$^{-1}$ (\citealt{fes+07}) to be the typical ejecta velocity, we find the total ejecta mass to be $\sim0.2$ M$_\odot$ and the kinetic energy is of $2.2\times10^{50}$ erg (taking a similar rise time as SN 2002bj of $\sim6$ days); this is consistent with our analysis of the SNR in section \ref{sub:Supernova-remnant}, showing that the energetics and ejecta mass are likely below those of canonical type Ia SNe.  
We conclude that all three SNe seem to have a low ejecta mass, $0.1-0.3$ M$_\odot$ and low kinetic energies, $1-3\times10^{50}$ erg.  

\subsection{Environments}

Both SNe 1885A and 1939B are found in old environments. Multi-band photometric observations of the environment of SN 1885A \citep{ols+06} in the bulge of M31 found it to be composed of an old stellar population, with a characteristic age of $\sim10$ Gyr.
SN 1939B occurred in the elliptical galaxy NGC 4621.  HST images of the galaxy
(WF4 detector, with the F555W and F814W filters and an exposure time
of 1050 s, taken on Feb 5$^{th}$, 1995)
and of the location of SN 1939B were used by  \citet{van+99} 
for PSF fitting and subtraction of foreground stars. Their analysis
yielded color measurement of $F555W-F841W=1.37$ mag for the unresolved
light within their error circle ($10''$), which was found to be consistent
with a background of K-giant stars. This confirms
that the close environment of SN 1939B shows no traces of star formation,
and likely excludes local star-formation activity in this elliptical galaxy at the SN position \citep{van+99}.
SN 2002bj is found at the outer region of a nearly face-on barred spiral galaxy (NGC 1821;
\citealt{poz+10}). As mentioned by \citet{poz+10}, the diverse stellar
population of such galaxies preclude any strong inference on the progenitor
system.

We note that the old environment found
for two out of the three SNe in the sample (where the properties of
the close environment of third, SN 2002bj, are not known) 
suggest the progenitors for these SNe are old, possibly older than those of regular type Ia SNe.  This may be similar to the case of SN 1991bg-like and SN 2005E-like events; 70 \% and 50 \% of which, respectively, are found in early type galaxies (E and S0; \citealp{how01,per+10}). The progenitors of these SNe are therefore most likely to be WDs.

\subsection{Rates}
Many calculations of SN rates have been reported in the literature, some
of which included the SNe discussed in this paper. Although these studies did not discuss the fast-evolving SNe specifically (except for those using
the LOSS survey data; \citealp{li+10,poz+10}), they can be used to
estimate how common these objects are. In table 1 we summarize
the relative fraction of such SNe, based on various SN rate estimates.
Poisson statistics are used to derive uncertainties.
Based on these calculations, we conclude that the various estimates
are generally consistent with a fraction of $1-2$ \% of all
observed SNe being bright and fast-evolving SNe.  Given the typically much
shorter timescale of these SNe (2-3 shorter than typical type Ia SNe), the actual fraction could be twice as large for large cadence surveys, and therefore our estimate likely serves as a lower limit, with an upper limit of  $\sim4-5$ $\%$.

\begin{table}
\label{rates}
\caption{Fraction rate of bright fast evolving SNe}
\begin{tabular}{|c|c|c|c|}
\hline 
Fast evolving & Total & Fraction (\%) & Ref.\tabularnewline
\hline
\hline 
1 (1939B) & 5 & $20\pm20$ & \citealp{zwi42}\tabularnewline
\hline 
1 (1939B) & 30 & $3.3\pm3.3$ & \citealp{kat+67}\tabularnewline
\hline 
1 (1885A) & 45$^*1$ & $2.2\pm2.2$ & \citealp{tam70}\tabularnewline
\hline 
2 (1885A, 1939B) & 96 & $2.1\pm1.4$ & \citealp{tam+94}\tabularnewline
\hline 
0 & 110 & $<0.9\pm0.9$ & \citealp{cap+97}\tabularnewline
\hline 
1 (2002bj) & 103$^**$ & $1\pm1$ & \citealp{poz+10}\tabularnewline
\hline
\end{tabular}

$^{*}$The sample is only of Sb and Sc galaxies

$^{**}$They reported SN 2002bj relative to 31 type Ia SNe, the total
number of SNe, 103, is estimated given the fractional rate of type
Ia SNe to be $\sim0.3$ \citep{li+10}. 
\end{table}
\section{Summary}

In this study we explored two fast evolving SNe:
1885A and 1939B. These SNe and the recently studied SN 2002bj display similar light curve behavior; they all show a fast rise to a bright peak magnitude followed by a rapid decay. These results point to a low mass ($<$M$_{\odot}$) of ejected material, which is consistent with our analysis of the SN 1885A remnant. This analysis
 shows the SNR to be highly peculiar and inconsistent with the typical
kinematics and ejecta masses of type Ia SNe. 
The spectra of the fast-evolving SNe are puzzling.  Our newly analyzed spectrum of SN 1939B shows no clear evidence for helium, and the spectra of SN 1885A likely does not show evidence for helium either. Note, however, that the spectrum of SN 1939B is of low quality, and a modern spectrum of SN 1885A is not available. Helium lines are possibly observed in the spectrum of SN 2002bj [but see Kasliwal et al. (2010), for a possible alternative interpretation]. 
None of the spectral data show evidence for hydrogen.
The environments of SNe
1885A and 1939B are old, in the bulge of M31 and in the elliptical
galaxy NGC 4621. The local environments of these SNe show no traces
of star-formation activity.  

The fast light curve and the low mass of ejected material of all three events 
suggest an emerging class of fast evolving type I
SNe, although more data are still required to form a clear observational
picture. If these events are related, we may conclude
that fast-evolving SNe are more likely to originate in old environments
where no star formation occurs, and are likely to result 
from the explosion of WDs. 
However, the specific explosion model for these SNe, and the consistency with suggested models (e.g. the helium-shell detonation scenario) are not yet known. 
Specifically, the role of helium, if any,
in these events, as well as the origin of the light curve differences between them (especially at late times) are unclear.  Modern spectroscopic follow-up of a larger sample of such objects is required to resolve these issues. Such objects can be found by surveys of large volumes at high cadence such as the Palomar Transient Factory (PTF) survey. Indeed, recently, such SN event (SN 2010X) have been found by PTF. Although this This event is found to be fainter than the fast SNe discussed here ($\sim-17$ mag in the r-band;  $\sim1.5$  mag fainter than SN 2002bj), it similarly shows fast rise and decay times ($\sim5$ days). A detailed discussion and analysis of this SN can be found in \citet{kas+10}.  

\acknowledgements{We thank Steve Reynolds, Zhiyuan Li, Mike Garcia and Eduardo Bravo for helpful discussions and data on SNR 1885A. We thank Dovi Poznanski and Ryan Chornock for helpful discussions about fast SNe and data on SN 2002bj. 
We also thank John Grula, George Carlson, Wal
Sargent, Hy Spinrad, Ivan King, Donna Kirkpatrick, Francois Schweizer,
George Preston, Andy McWilliam, Doug Mink and Bob Kirshner for helping with locating, reducing and analyzing of the original spectroscopic plate
of SN 1939B. We would also like to thank the anonymous referee, and the second referee, David Branch, for their helpful comments. HBP is a CfA and BIKURA(FIRST) fellow. 
A.G. is supported by grants from the Israeli Science Foundation,
an EU FP7 Marie Curie IRG Fellowship, a research grant from the
Peter and Patricia Gruber Awards, the Weizmann Minerva program and
the Benoziyo Center for Astrophysics.}

\bibliographystyle{apj}

\end{document}